\documentclass[aps,prl,superscriptaddress,reprint,showpacs,twocolumn,
preprintnumbers,amsmath,amssymb,floatfix]{revtex4-1}
\usepackage{graphicx}
\usepackage{epstopdf}
\usepackage[latin1]{inputenc}
\usepackage{amssymb}
\usepackage{bm}
\usepackage{stmaryrd}
\usepackage{wasysym}
\usepackage{subfigure}
\usepackage{afterpage}

\newcommand{\asnmr}{$^{75}$As }

\newcommand{\lafemnasof}{LaFe$_{1-x}$Mn$_{x}$AsO$_{0.89}$F$_{0.11}$ }

\newcommand{\mossbauer}{M\"{o}ssbauer }

\begin{document}

\title{Fast recovery of the stripe magnetic order by Mn/Fe
substitution \\ in F-doped LaFeAsO superconductors}
\author{M. Moroni}
\email{matteo.moroni01@universitadipavia.it}
\affiliation{Department of Physics, University of Pavia-CNISM, I-27100 Pavia,
Italy}
\author{P. Carretta}
\affiliation{Department of Physics, University of Pavia-CNISM, I-27100 Pavia,
Italy}
\author{G. Allodi}
\affiliation{Dipartimento di Fisica e Scienze della Terra, Universit\`{a} di
Parma,  I-43124 Parma, Italy}
\author{R. De Renzi}
\affiliation{Dipartimento di Fisica e Scienze della Terra, Universit\`{a} di
Parma,  I-43124 Parma, Italy}
\author{M. N. Gastiasoro}
\affiliation{Niels Bohr Institute, University of Copenhagen, Juliane Maries Vej
30, 2100 Copenhagen, Denmark}
\author{B. M. Andersen}
\affiliation{Niels Bohr Institute, University of Copenhagen, Juliane Maries Vej
30, 2100 Copenhagen, Denmark}
\author{P. Materne}
\affiliation{Institute of Solid State Physics, TU Dresden, D-01069 Dresden,
Germany}
\author{H.-H. Klauss}
\affiliation{Institute of Solid State Physics, TU Dresden, D-01069 Dresden,
Germany}
\author{Y. Kobayashi}
\affiliation{Department of Physics, Division of Material Sciences, Nagoya
University, Furo-cho, Chikusa-ku, Nagoya 464-8602, Japan}
\author{M. Sato}
\affiliation{Department of Physics, Division of Material Sciences, Nagoya
University, Furo-cho, Chikusa-ku, Nagoya 464-8602, Japan}
\author{S. Sanna}
\affiliation{Department of Physics, University of Pavia-CNISM, I-27100 Pavia,
Italy}
\affiliation{Department of Physics and Astronomy, University of Bologna, 40127
Bologna, Italy}

\date{\today}

\begin{abstract}

$^{75}$As Nuclear Magnetic (NMR) and Quadrupolar (NQR) Resonance
were used, together with M\"{o}ssbauer spectroscopy, to
investigate the magnetic state induced by Mn for Fe substitutions
in F-doped LaFe$_{1-x}$Mn$_{x}$AsO
superconductors. The results show that $0.5$\% of Mn doping is enough to
suppress the
superconducting transition temperature $T_c$ from  27 K to zero and to
recover the magnetic structure observed in the parent undoped
LaFeAsO. Also the tetragonal to orthorhombic transition
of the parent compound is recovered by introducing Mn, as
evidenced by a sharp drop of the NQR frequency. The NQR spectra also show that
a charge localization process is at play in the system.
Theoretical calculations using a realistic five-band model show
that correlation-enhanced RKKY exchange interactions between nearby
Mn ions stabilize the observed stripe magnetic order.
These results give compelling evidence that F-doped LaFeAsO is a
strongly correlated electron system at the verge of an electronic instability.

\end{abstract}


\maketitle


The interplay between impurity induced disorder and electronic
correlations often gives rise to complex phase diagrams in
condensed matter~\cite{miranda2005, lee1985}. The electronic
correlations drive a system towards a quantum phase
transition, as it is typically found in the fullerides~\cite{capone2015} and in
heavy-fermion compounds~\cite{curro2014, curro2016}, with an
enhancement of the local susceptibility and, hence, a small
perturbation, as the one associated with a tiny amount of
impurities, can significantly affect the electronic ground-state
~\cite{Alloul2009,sato2012, sanna2013, prando2015}. In
the cuprates and in the electron-doped iron-based superconductors
(IBS) the strength of the electronic correlations can be tuned
either by charge doping or by applying an external or a chemical
pressure.\cite{luetkens2009, sanna2009, shiroka2011, prando2013,
khasanov2011} In particular, upon increasing the charge doping, the
strength of the electronic correlations tend to decrease
~\cite{Capone2002, hess2009, Qazilbash2009, Ikeda2010, Dai2012,
Yu2013, Medici2014, Lee2010} and a metallic Fermi liquid (FL)
ground state is usually restored~\cite{mazin2008, tropeano2010,
giovannetti2011, benfatto2011, gastiasoro2016}. However
significant electronic correlations may still be present even
close to the charge doping levels yielding the maximum
superconducting transition temperature $T_c$ and a convenient
method to test their magnitude is to perturb the system with
impurities.

\begin{figure}[h!]
\includegraphics[width=8.6cm,keepaspectratio]{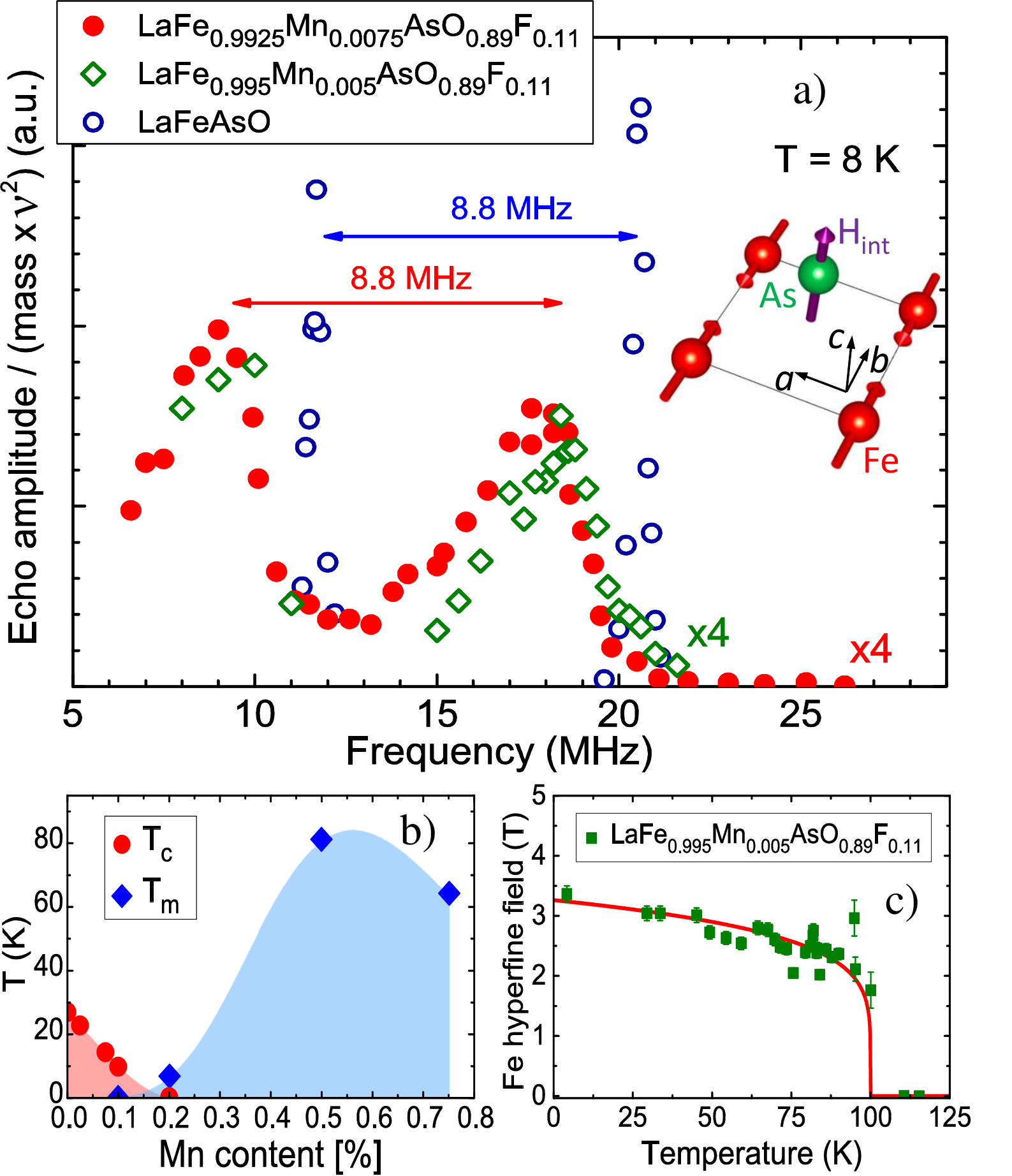}
\caption{a) Comparison of
the \asnmr Zero-Field NMR spectra between the LaFeAsO
parent compound and \lafemnasof for $x=0.5$ \% (green) and 0.75 \% (red). For
the sake of comparison the
intensity of the spectra for the Mn-substituted compounds are multiplied by 4.
Inset: sketch of the
magnetic unit cell for the stripe order (the red arrows represent the Fe
magnetic
moments directions while the magenta arrow corresponds to the orientation of
the internal field at the As site) b)
Electronic phase diagram of LaFe$_{1-x}$Mn$_{x}$AsO$_{0.89}$F$_{0.11}$. T$_c$
and T$_m$ were determined from magnetization (SQUID) and zero
field $\mu$SR measurements, respectively (see Ref.~\onlinecite{Hammerath2014}).
c)
Temperature dependence of the hyperfine magnetic field, $B_h$, for $x=0.5$ \% as
derived from \mossbauer spectra. The red solid line tracking the
order parameter is a phenomenological fit of $B_h$ with $B_h =
3.3(1-(\textrm{T}/\textrm{T}_m))^\beta$ where T$_m = 100$ K and $\beta =
0.17$.} \label{fig:zf-nmr}
\end{figure}

The introduction of Mn impurities at the Fe sites was reported to
strongly suppress $T_c$ in several IBS, both of the
BaFe$_2$As$_2$~\cite{cheng2010,li2012,leboeuf2013} and of the
LnFeAsO (Ln1111, Ln=Lanthanides)~\cite{sato2010} families.
Within the Ln1111 family the effect of impurities is
particularly significant in
La1111~\cite{gastiasoro2016,Hammerath2014}. 
In fact, while in most the IBS compounds the T$_c$ suppression rate
(d$T_c/$dx) is well below~10~\mbox{K/\% Mn}, in
LaFeAsO$_{0.89}$F$_{0.11}$ just 0.2 - 0.3\% of Mn impurities
suppress superconductivity from the optimal T$_c\simeq 27$ K
(d$T_c/$dx $\sim$~110~\mbox{K/\% Mn}) and then, at higher Mn
doping levels, a magnetic order develops (see
Fig.~\ref{fig:zf-nmr}b)~\cite{Hammerath2014, hammerath2015,
moroni2016}. The understanding of
why such a dramatic effect is present, what type of magnetic order
is developing and how to describe these materials at the
microscopic level are presently subject of
debate~\cite{fernandes2013,tucker2012,gastiasoro2015}. Here we
show, by means of zero-field (ZF) NMR, nuclear quadrupole
resonance (NQR) and M\"ossbauer spectroscopy that the introduction
of 0.5 \% of Mn in LaFeAsO$_{0.89}$F$_{0.11}$ induces the recovery
of the magnetic order and of the tetragonal to orthorhombic (T-O)
structural transitions observed in LaFeAsO, the parent compound of
La1111 superconductors. Moreover the decrease of the charge
transfer integral and the enhanced electron correlations lead to
the electron localization and to a local charge distribution
similar to that found in LaFeAsO. We also present
theoretical calculations showing that correlation-enhanced RKKY
exchange couplings between neighboring Mn ions stabilize the
magnetic order characterized by $Q_1=(\pi,0)$ and $Q_2=(0,\pi)$
domains.

The \lafemnasof polycrystalline samples under investigation are the same
ones studied in Ref.~\onlinecite{Hammerath2014}. Further details
on the sample preparation and characterization can be found in the Suppl.
Materials~\cite{suppmat}.
$^{75}$As zero-field (ZF) NMR spectra were obtained by recording
the echo amplitude as a function of the irradiating frequency in
the 6-26 MHz range for $T=8$ K (see Fig.~\ref{fig:zf-nmr}a). Both the spectra of the LaFeAsO parent compound and of the x=0.75\%
sample are characterized by two peaks, which in the latter compound are
rigidly shifted to lower frequencies and broadened. The
peaks are associated with the $m_I=1/2\rightarrow m_I=-1/2$ and
$m_I=-1/2\rightarrow m_I=-3/2$ transitions, with $m_I$ the
component of the nuclear spin $I$ along the quantization axis,
which in the case of a stripe magnetic order (magnetic wave vector $Q=(\pi,0)$
or $(0,\pi)$), as it is the case
for LaFeAsO, is along the $c$ axis. The frequency shift between
the two peaks is given by the nuclear quadrupole frequency
determined by the local charge distribution, which
at 8 K is $\nu_Q= 8.8$ MHz both for LaFeAsO and for the $x=0.75$
\% sample.

The position of the low-frequency peak ($\nu_c$), associated with
the $1/2\rightarrow -1/2$ transition, is determined by the
magnitude of the hyperfine field at $^{75}$As, and one can write
that $\nu_c= (\gamma/2\pi) |\mathcal{A}\langle\vec S \rangle|$,
with $\gamma$ the $^{75}$As gyromagnetic ratio, $\mathcal{A}$ the
hyperfine coupling tensor and $\langle\vec S \rangle$ the average
electron spin, corresponding to the magnetic phase order parameter.
Accordingly, the low-frequency shift of the two
peaks in the sample with $x=0.75$ \% would indicate a reduction of
the order parameter to about 80\% of the value found for LaFeAsO.
The sample with $x=0.5$ \% displays a very similar behavior with a
slight increase in the low-temperature order parameter, following
its slightly higher magnetic transition temperature
(T$_m$)~\cite{Hammerath2014}. From the magnetic point of view the
two samples $x=0.5$ \% and 0.75 \% are almost equivalent, as
already shown from previous muon spin relaxation
experiments~\cite{Hammerath2014}.

In order to further study the magnetic order parameter  we
measured the temperature dependence of M\"ossbauer spectra for the
$x=0.5$ \% sample. Fig.~\ref{fig:zf-nmr}c shows that in the low
temperature limit the internal field at the Fe site is of about
3.5 T, i.e. the magnitude of the order parameter is reduced to
about 70\% of the value found in pure
LaFeAsO~\cite{klauss2008,kitao2008,mcguire2008}, in reasonable
agreement with what we derived above from ZF-NMR.

Now, one has first to consider if magnetic
orders different from the stripe one could give rise to a similar ZF-NMR
spectrum, taking into
account the reduction in the Fe moment to about 80\% of the value
found in LaFeAsO. The other low-energy magnetic orders which could develop in
this compound are the N\'eel ($Q=(\pi,\pi)$) and the Orthomagnetic type,
with a $\pi/2$ rotation of the adjacent spins~\cite{giovannetti2011}.
Calculations of the hyperfine magnetic
field~\cite{kitagawa2008,kitagawa2010} at the As site (Suppl. Materials~\cite{suppmat}) show that
both these magnetic orders would give rise to ZF-NMR lines
significantly shifted from the ones reported in
Fig.~\ref{fig:zf-nmr}a, thus confirming that the order is stripe-type. On the other hand, one could argue that the stripe order
could coexist with other types of order developing close to Mn
impurities and that we are actually detecting the signal from a
fraction of $^{75}$As nuclei only. Thus, we have performed a
quantitative estimate of the amount of nuclei contributing to the
$x=0.75$~\% sample ZF-NMR spectrum in Fig.~\ref{fig:zf-nmr}a by comparing its integrated intensity with
that of the LaFeAsO sample, where 100 \% of the sample is in the
stripe collinear phase. We found that $95\pm 5$ \% of the
$x=0.75$\% sample is in the stripe order.

\begin{figure}[t]
\includegraphics[height=6cm,
keepaspectratio]{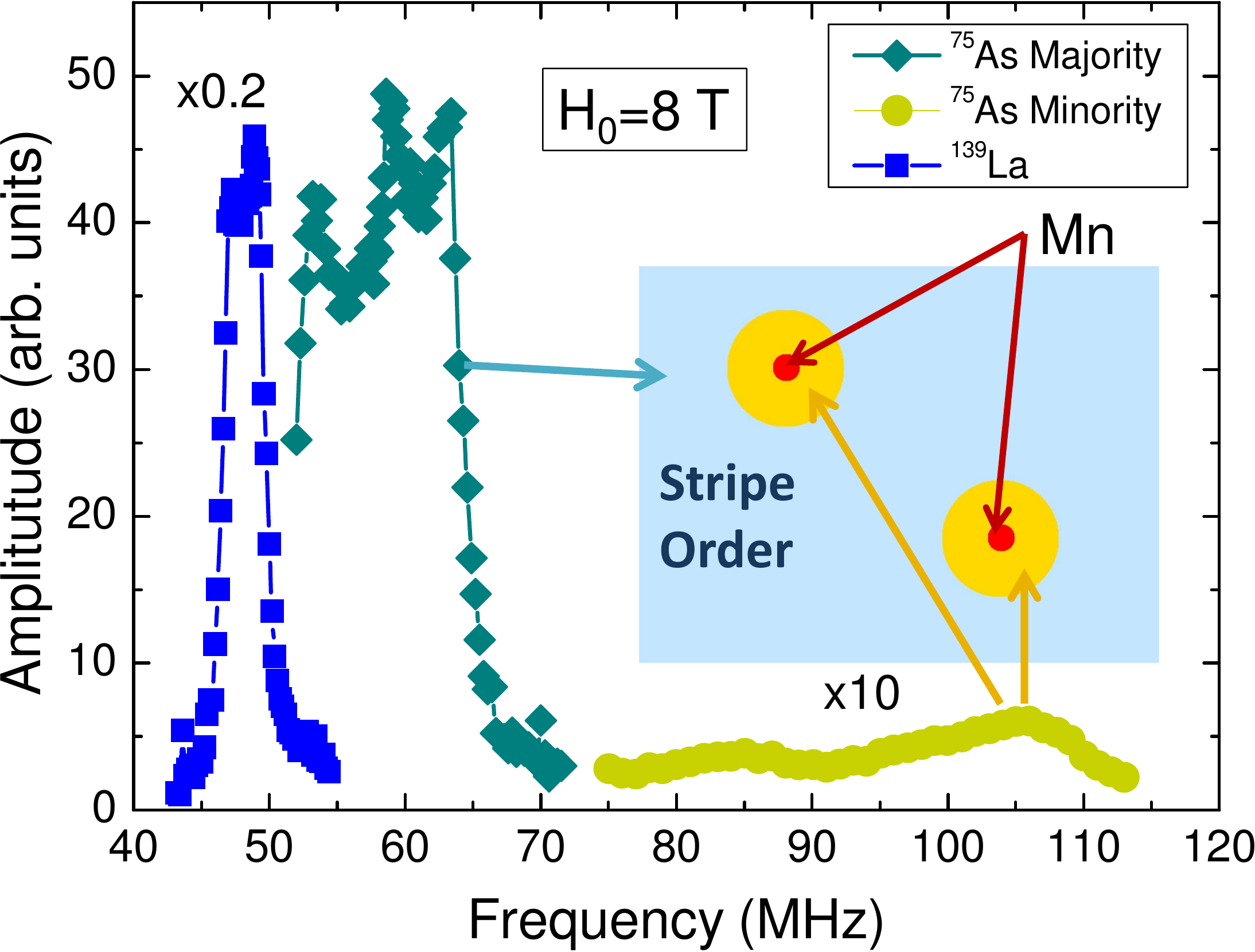} \caption{\lafemnasof ($x=0.5$\%) $^{75}$As and
$^{139}$La NMR spectra in the 40-115 MHz frequency range, measured
at 10 K, for
an applied magnetic field $\mu_0$H = 8 T. Inset: pictorial
representation of the Fe layer for few per-thousand of Mn
substitution in LaFeAsO$_{0.89}$F$_{0.11}$} \label{fig:NMRspectrum}
\end{figure}

In order to further check if there is a small ($\leq 5$ \%)
fraction of $^{75}$As nuclei that we are missing, we performed
$^{75}$As NMR measurements in a 8 T magnetic field. In
Fig.~\ref{fig:NMRspectrum} the powder NMR spectrum displays a
large fraction of nuclei with a spectrum broadened~\cite{riedi1989} by the internal
field developing in the stripe phase (cyan diamonds) as well as a
small fraction of about $3\pm 1$ \% of $^{75}$As nuclei with a
significant NMR shift towards higher frequencies (yellow circles).
These latter nuclei are likely the ones close to Mn impurities
where a large hyperfine field is expected. For $x=0.5$\% there are
2\% of As nuclei which are nearest neighbors of a Mn impurities, a
value very similar to the one we found. Hence, the introduction of
Mn suppresses superconductivity and leads to the recovery of the
stripe magnetic order found in the parent LaFeAsO compound. Any
incommensurate magnetic order, if present, should have a magnetic
wave-vector very close to the stripe one (see Suppl.
Material~\cite{suppmat}). This aspect could be clarified by future neutron diffraction experiments.

\begin{figure}[b]
\includegraphics[height=5.5cm,
keepaspectratio]{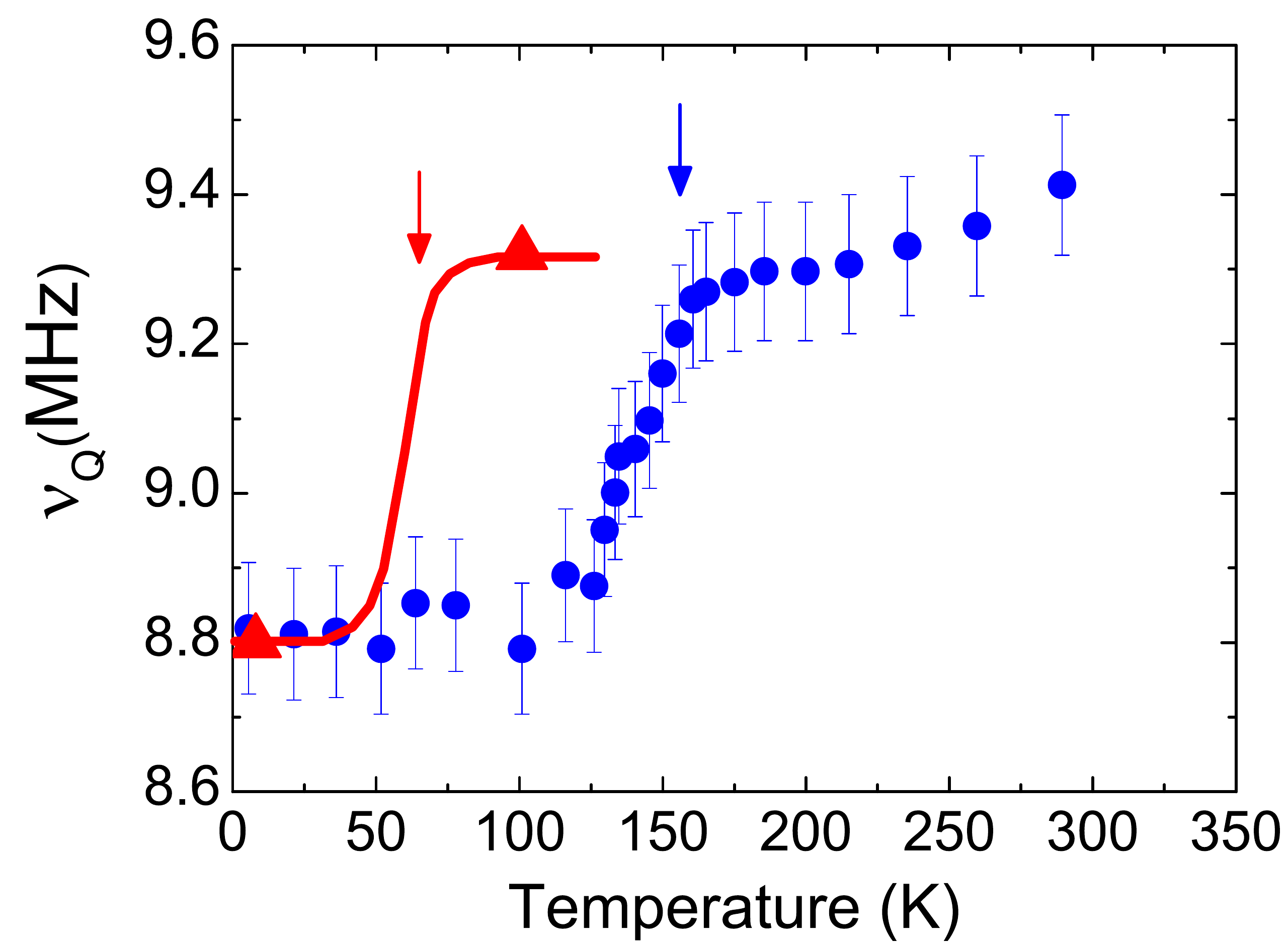} \caption{Quadrupolar
frequency, $\nu_Q$, as a function of temperature for the LaFeAsO
(blue circles, data from Ref.\cite{fu2012}) and \lafemnasof with
$x=0.75$ \% (red triangles). For the latter the point above $T_m$
is taken as the frequency of the low-frequency peak of
Fig.~\ref{fig:nqrspectra}a (see text).
The vertical arrows indicate the magnetic
transitions, while the red solid line is a guide to the eye.}
\label{fig:structransition}
\end{figure}
\begin{figure}[t]
\includegraphics[width=8.6cm,
keepaspectratio]{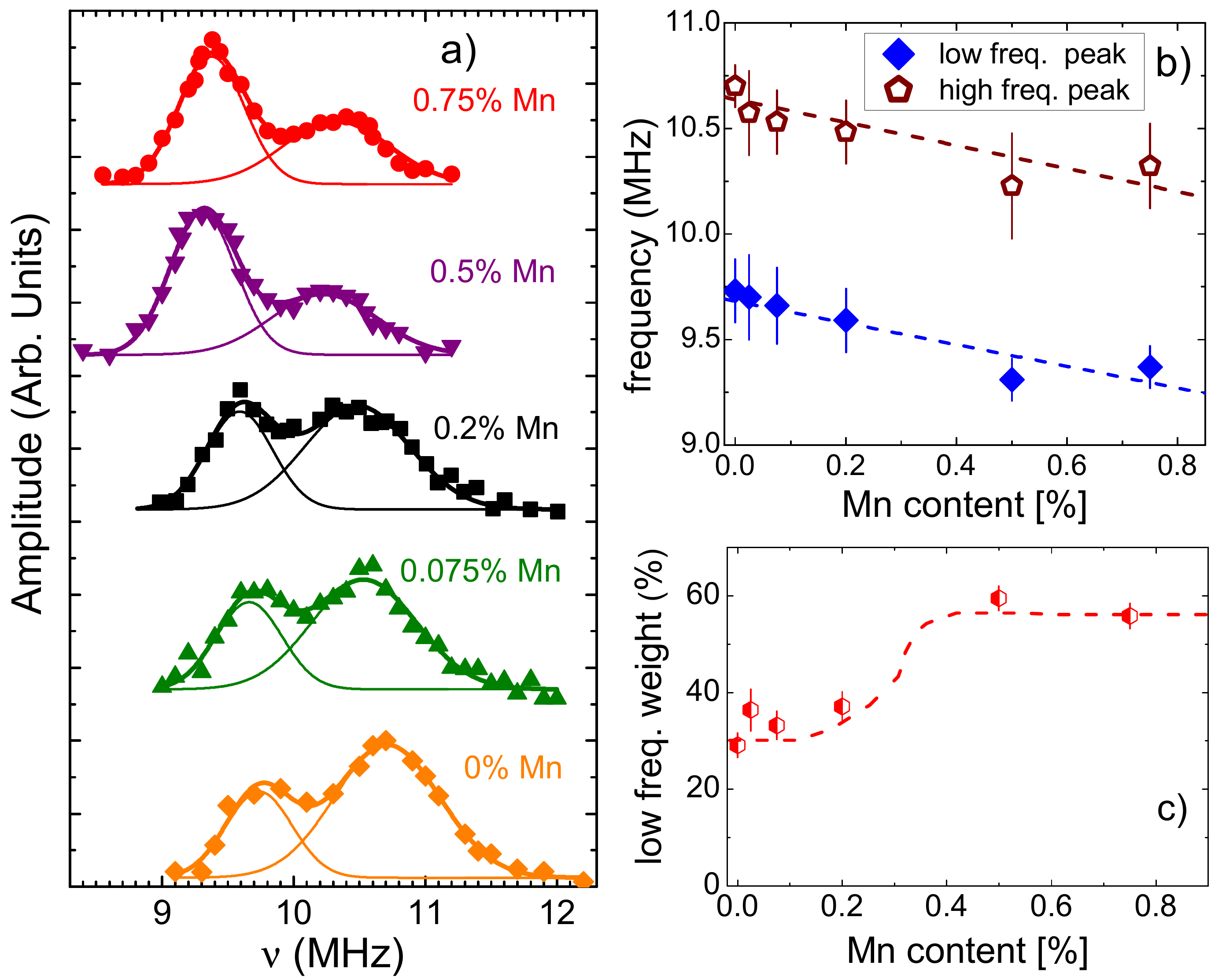} \caption{ a) $^{75}$As NQR
spectra of \lafemnasof for different Mn contents,
measured at $T= 77$ K for $x\leq 0.2$~\% and $T=100$ K (above
T$_m$) for $x>0.2$ \%. The solid lines are fits to a sum of two Gaussian
functions. b)
Frequency of the low (blue diamonds) and high energy (red pentagons) peaks as a
function of Mn content. The dashed lines are a guide to the eye. c) Weight of
the low frequency peak as a function of Mn content, the dashed line is a guide
to the eye.}
\label{fig:nqrspectra}
\end{figure}

The nuclear
quadrupole frequency (Fig.~\ref{fig:structransition}) shows a jump on passing
from just above T$_m$ ($^{75}$As NQR) to below T$_m$ ($^{75}$As ZF-NMR) which
is very similar to the one detected \cite{fu2012} in LaFeAsO.
This abrupt change in $\nu_Q$ is associated with the T-O
distortion. Therefore, the observation of a similar change in
$\nu_Q$ for the $x=0.5$ \% compound indicates that when the stripe
magnetic order is recovered by Mn doping also the T-O structural
transition is recovered, confirming that this transition is closely
related to the onset of large stripe magnetic correlations. We further
remark that the T-O transition causes also a change in the
electric field gradient probed by Fe nuclei, as detected by M\"ossbauer
spectroscopy (see Suppl. Materials).

Another relevant aspect can be grasped by looking at the $^{75}$As
NQR spectra which can give detailed informations about the
electronic environment surrounding the As nuclei~\cite{lang2016,lang2010,kobayashi2010}.  The spectra in
Fig.~\ref{fig:nqrspectra},
measured at $T= 77$ K
for $x\leq 0.2$~\% and at $T=100$ K (above T$_m$) for $x>0.2$ show a clear shift of the NQR spectrum towards lower
frequency with increasing Mn content (see
Fig.~\ref{fig:nqrspectra}b) and a rapid change in the intensity of
the low-frequency peak for $x>0.2$ \%. It is worth to
note that for $x>0.2$ \%
the frequency of the dominant low frequency peak perfectly matches
that of the paramagnetic phase of LaFeAsO (see
Fig.~\ref{fig:structransition} and~\ref{fig:nqrspectra}b),
indicating a similar electronic ground state. According to Lang et
al. \cite{lang2010}, the low and high-frequency NQR peaks should
be associated with nanoscopically segregated regions with
different electron doping levels. In particular, the low-frequency
peak should be associated with a lower electronic concentration of
weakly itinerant electrons. Hence, the increase in the magnitude
of the low-frequency peak above $x=0.2$~\% indicates a tendency
towards electron localization. This finding is also corroborated
by the rapid increase of the electric resistivity as a function of
Mn content previously observed \cite{sato2010,berardan2009} across the
metal-insulator crossover taking place around $x= 0.2$\%. A
similar rise in resistivity was also observed \cite{hess2009} in
LaFeAsO$_{1-x}$F$_{x}$ with decreasing F content. 
One would expect that since LaFeAsO$_{1-x}$F$_{x}$ is characterized by 
Fermi pockets, a
scattering center as Mn would induce in any way charge localization. However,
this can occur only if the response function of the bare system is strongly
enhanced, as it is the case for La1111 but not for Sm1111.
We add here that the increase of the resistivity and the slight
increase of the lattice
constant $c$ follow the suppression of the superconductivity (see Fig. 12 of
Ref.~\cite{sato2010} for details). In fact, the $c$ axis value appears to
correlate with the T$_c$ value, irrespective
of the microscopic mechanism of suppression.

In order to further clarify the origin of the dramatic effect of
Mn doping in LaFeAsO$_{0.89}$F$_{0.11}$, we carried out real space
theoretical calculations using a realistic five band Hamiltonian~\cite{ikeda2010,gastiasoro2016} (see Supplementary Material~\cite{suppmat}).
In Ref.~\cite{gastiasoro2016} it was demonstrated that in this
framework the enhanced spin correlations developing around Mn
severely speed up the reduction of $T_c$ driven by the magnetic
disorder, and may quench the entire superconducting phase already
at Mn concentrations below 1 \%. The Mn moments, while
substituting random Fe positions, orient their moments favorably
to generate a long-range ordered SDW phase which minimizes the
total free energy of systems at the brink of a SDW
instability.\cite{andersen2007,gastiasoro2014}

\begin{figure}[h!]
\includegraphics[height=7cm,keepaspectratio]{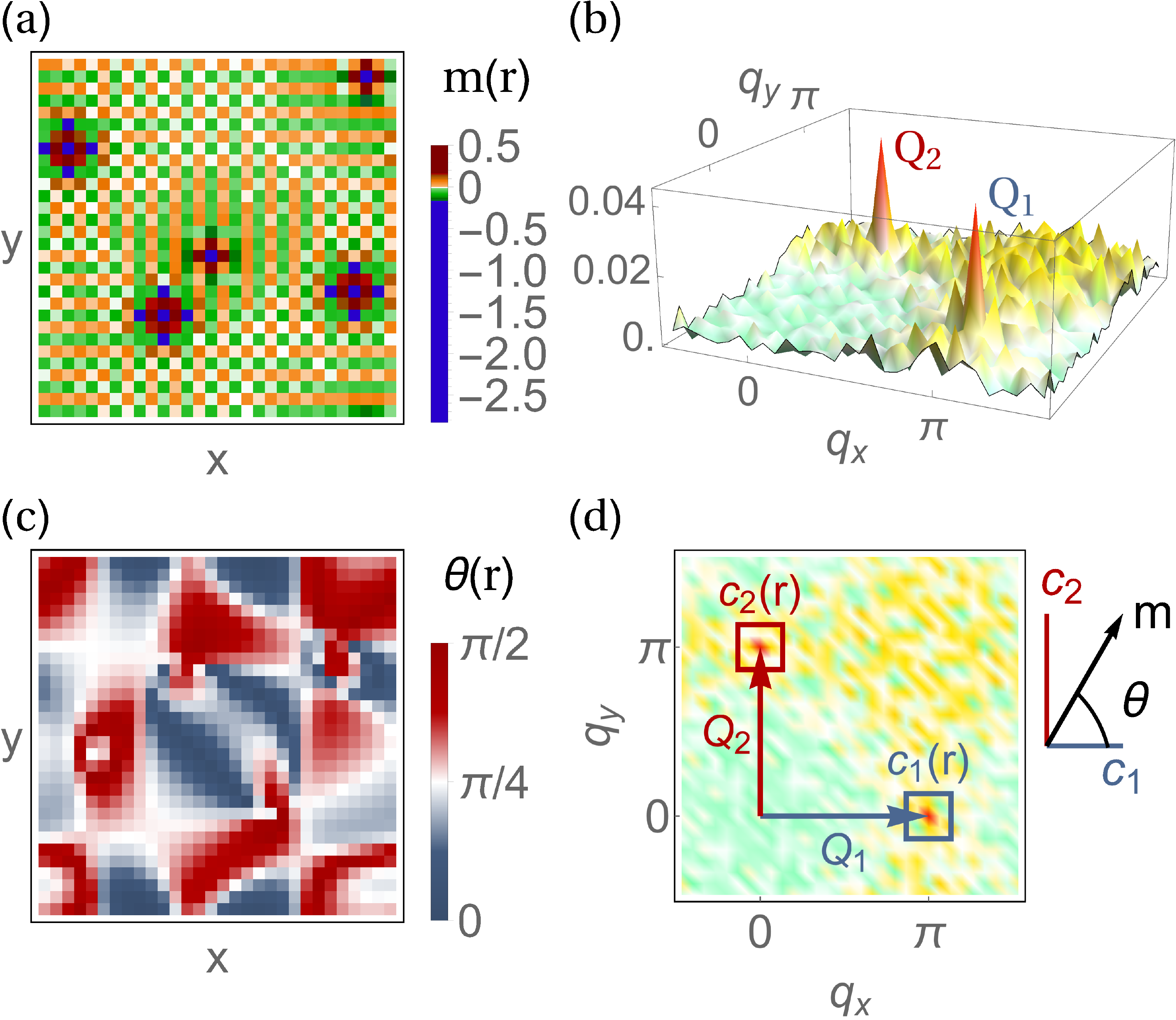}
\caption{(a) Induced magnetic order $m(r)$ for 0.55\% Mn moments and (b) its
Fourier transform $|m(q)|$.
(c) Local phase map $\theta(r)=\arctan(|c_1(r)|/|c_2(r)|)$ showing the ordering
vectors on the lattice: $\theta=0$ and $\pi/2$ for single-Q $Q_1$ (blue) and
$Q_2$ (red) domains respectively, and $\theta=\pi/4$ for double-Q regions
(white).
The coefficient associated with $Q_1$ ($Q_2$), $c_l(r)=\sum_n c_n \exp
[i(q_n-Q_l)r]$, is calculated by a filtered Fourier transform with the
$\{q_n\}$ wave vectors contained inside the blue and red squares shown in panel
(d).
The inset illustrates the definition of the local phase $\theta (r)$.}
\label{fig:theory}
\end{figure}

In Fig. 5(a) we show the total magnetization for a collection of
0.55\% Mn ions randomly placed in the square Fe lattice. This
concentration of Mn is able to fully suppress $T_c$ and spin
polarize all Fe sites (which were all non-magnetic without Mn
impurity ions). The Mn-induced magnetic order existing in the
inter-impurity regions is long-ranged as reproduced by the sharp
peaks in Fig.~5(b), which are absent in the Mn free compound. A small fraction
of the sites, roughly
corresponding to the Mn sites and to their nearest-neighbors
(amounting to~$\sim 5\%$ of the lattice) exhibits a significantly
larger moment, in overall agreement with the above discussion of
the $^{75}$As NMR data (see Fig.~\ref{fig:NMRspectrum}).

The magnetic order generated by Mn doping is efficiently
stabilized due to correlation-enhanced RKKY exchange couplings
between neighboring Mn ions. The structure of the induced order is
thus dictated by the susceptibility of the bulk itinerant system
which, in the present case, is peaked at $Q_1=(\pi,0)$ and
$Q_2=(0,\pi)$ regions. In Fig.~5(c) we provide a real-space map of
the dominant momentum structure by utilizing a filtered Fourier
transform illustrated in Fig.~5(d) and the associated caption. As
seen, the system breaks up into regions of single-Q domains, i.e.
either $Q_1$- or $Q_2$-dominated regions, and does not exhibit
substantial volume fraction of double-Q
order.\cite{gastiasoro2015} This is consistent with the presence
of a (reduced) orthorhombic transition associated with the
Mn-induced magnetic order, as found by $^{75}$As NQR
(see Fig.~\ref{fig:structransition}).
All these theoretical results match with the experimental
outcomes.

Overall the above scenario is a clear indication that in
LaFeAsO$_{0.89}$F$_{0.11}$ superconductivity emerges from a
strongly correlated electron system close to a metal-insulator
transition. The electron correlations are so strong that, owing to
the enhanced spin susceptibility at $Q=(\pi,0)$, the effect of a
tiny amount of impurities extends over many lattice sites, giving
rise to a sizable RKKY coupling among them able to abruptly
destroy superconductivity and to restore the stripe magnetic
order. The onset of the magnetic order is intimately
related with the charge localization~\cite{sato2010} and hence to
the T$_c$ suppression. This situation is reminiscent of the
phenomenology observed in heavy fermion (HF)
compounds~\cite{doniach77,weng-rev2016} where the FL phase
vanishes and a magnetic order arises in correspondence of a QCP
when the RKKY interaction overcomes the Kondo coupling. For
example, in CeCoIn$_5$, a HF compound, it was shown that a tiny
amount of Cd doping restores the long range antiferromagnetic (AF)
order~\cite{curro2014,curro2016,pham2006} and suppresses the
superconducting dome developing around the quantum critical point
separating the FL from the AF phase. In \lafemnasof the Kondo
coupling would involve itinerant and more localized 3$d$
electrons~\cite{wu2016,haul2009,dai2012} playing a role analogous
to the $f$ electrons in the HF. When these latter
electrons finally localize the RKKY coupling among Mn impurities
leads to the recovery of magnetism and the suppression of the
metallic superconducting state.

The abrupt suppression of the superconducting phase and the
recovery of the magnetic order and of the structural T-O
transition give compelling evidence that the optimally F doped
LaFeAsO is at the verge of an electronic instability, close
to a QCP~\cite{Hammerath2014}. Previous experimental results have
shown that this system can be driven away from the
QCP via the total substitution of La with Nd or by the partial
substitution with Y~\cite{hammerath2015,moroni2016}, which shrink
the structure and cause a reduction of the electronic 
correlations~\cite{gastiasoro2016}. Hence, the Ln1111 compound can be
considered as a formidable example of how the electronic
properties of strongly correlated systems can be significantly affected by 
fine-tuning the correlation strength with impurities
and chemical pressure.

\begin{acknowledgments}
B. B\"{u}chner is thanked for useful discussions.  This work was supported by
MIUR-PRIN2012 Project No. 2012X3YFZ2. M.N.G. and B.M.A acknowledge support from
Lundbeckfond fellowship (grant A9318).
\end{acknowledgments}



	\newpage \mbox{ } \newpage
	
	\section{\textit{Supplementary Material for:}\\Fast recovery of the stripe magnetic order by Mn/Fe
		substitution  in F-doped LaFeAsO superconductors}

	\section{I. Sample preparation and characterization.}
	
	In the present study we have presented data for two series of
	samples (say, \#A and \#B) among many series showing a very similar trend of T$_c$ with Mn doping. The samples
	with the Mn concentration x = 0.5\% and 0.75\% belong to \#A, and
	x=0.075\% and 0.2\% belong to \#B. The x=0 sample displays an almost
	optimum value of T$_c$=27 K by SQUID magnetometry
	and muon spin rotation measurements, as reported in Ref.\cite{Hammerath2014s}. 
	Despite this fact, Fig.~4a shows that
	its high temperature NQR spectrum is characterized by
	a double peak structure, typically observed in the underdoped
	compounds \cite{lang2010s,kobayashi2010s} where charge-poor and
	charge-rich environments coexist at the nanoscale. Hence,
	following Ref. \cite{lang2010s}, the NQR results show that the
	whole set of samples displays intrinsic electronic inhomogeneity, 
	as pointed out for $x$=0-0.2 in Ref.\cite{Hammerath2014s}.
	The intensity of the $^{19}$F
	NMR signal measured at room temperature shows that the
	fluorine content is constant throughout all the samples within
	an accuracy of $\Delta y$ = 2\%. 
	
	%
	%
	
	\section{II. \mossbauer measurements}
	
\begin{figure}[ht]
		\includegraphics[width=8.5cm,
		keepaspectratio]{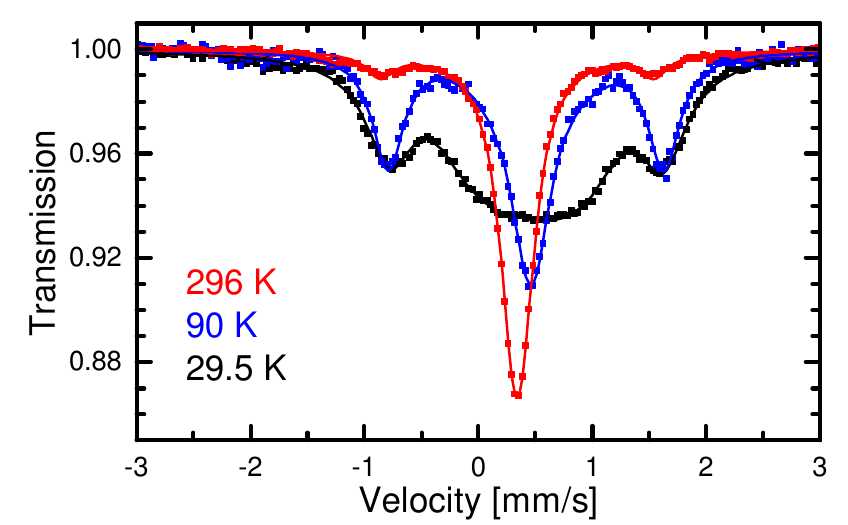} \caption{M\"ossbauer spectra for the $x=0.5$\% compound at a few selected temperatures both above and below T$_m$.}
		\label{fig:moss}
	\end{figure}

	The \mossbauer measurements were performed in transmission
	geometry in the 2.3 K - 300\;{}K temperature range using a CryoVac
	Konti IT cryostat on the \lafemnasof for x=0.5\%. As the $\gamma$
	source, a $^{57}$Co in rhodium matrix was used. We used Ferrocen
	powder to measure the influence of the experiment on the line
	width. The data was analyzed using the transmission integral.
	\mossbauer data for representative temperatures above and below
	T$_m$ are shown in Fig.~\ref{fig:moss}. At all investigated
	temperatures a three peak structure is observed. The two outer
	peaks at $\approx$\;{}-\;{}0.80 and 1.65\;{}mm/s correspond to the
	Ferrocen reference absorber while the inner peak is identified
	with the LaFe$_{1-x}$Mn$_{x}$AsO$_{0.89}$F$_{0.11}$.
	At 296\;{}K a non-resolved doublet structure is observed due to
	the interaction of the nucleus with an electric field gradient
	(EFG). In the principal-axis system, the EFG is fully determined
	by its \textit{z} component $V_{ZZ}$ and the asymmetry parameter
	$\eta = (V_{XX} - V_{YY})/V_{ZZ}$. We obtained a center shift of
	0.451(1)\;{}mm/s and a value of $V_{ZZ}$\;{}=\;{}4.5\;{}V/\AA$^2$.
	$\eta$ was found to be zero at all temperatures. The center shift
	increases upon cooling due to the temperature-dependent
	second-order Doppler effect to a value of 0.586(8)\;{}mm/s at
	4.2\;{}K. $V_{ZZ}$ increases with decreasing temperature to a
	value of 8.6\;{}V/\AA$^2$ at 82\;{}K. At the magnetic phase
	transition $V_{ZZ}$ increases to 12(2)~V/\AA$^2$ and remains
	constant within error bars down to lowest temperatures. This change of the the electric
	field gradient probed by Fe around the magnetic transition reflects the structural 
	\mbox{T-O} transition in agreement with the one probed 
	by As in the NQR measurements discussed in the main text.
	Below $T_m$ a broad peak is observed which is identified with a
	non-resolved sextet structure due to a field distribution. This
	magnetic field distribution was modeled using a Gaussian
	distribution. The temperature dependence of the first moment is
	shown in Fig.~1c of the main part of the paper.

	\section{III. NQR and NMR Spectra}
	In order to carry out the NMR experiments the polycrystalline \lafemnasof samples were crushed to a fine powder to improve radio frequency penetration.
	Since $^{75}$As is a spin I = 3/2 nucleus, above T$_m$, the NQR spectrum is characterized by a single line at a frequency
	\begin{equation} \label{hyperfine}
	\nu_Q= \dfrac{eQV_{ZZ}}{2h}\left(1+\dfrac{\eta^2}{3} \right)^{1/2}
	\end{equation}
	with $Q$ the electric quadrupole moment of the $^{75}$As nucleus,
	$V_{ZZ}$ the main component of the electric field gradient (EFG)
	tensor at the As site generated by the surrounding charge
	distribution and $\eta$ its asymmetry. Since LaFeAsO is tetragonal
	$\eta=0$ and $c$ is the quantization axis.  Thus, the broadening
	of the line is mainly due to the disorder present in the system
	since the EFG strongly depends on the local charge distribution.
	The two peak in the NQR spectra (Fig.4, main article) are due to an intrinsic electronic inhomogeneity (due to F doping) already present without Mn in superconducting samples where the whole volume is superconducting (see Ref.~\onlinecite{lang2010s}). 
	
	Below T$_m$, in case of a stripe magnetic order, an internal field
	H$_\mathrm{int} \parallel c$ is present at the As nuclei and we
	can perform standard NMR experiments with the only difference that
	the magnetic field is not provided by an external magnet but by
	the magnetic ordering of the Fe moments (see also the main part of
	the paper). Even if below T$_m$ the unit cell is orthorhombic the
	asymmetry $\eta$ is still small ($\eta\sim 0.15$) as reported in
	Ref.~\cite{fu2012s}.

	\begin{figure}[b]
		\includegraphics[width=6.2cm,
		keepaspectratio]{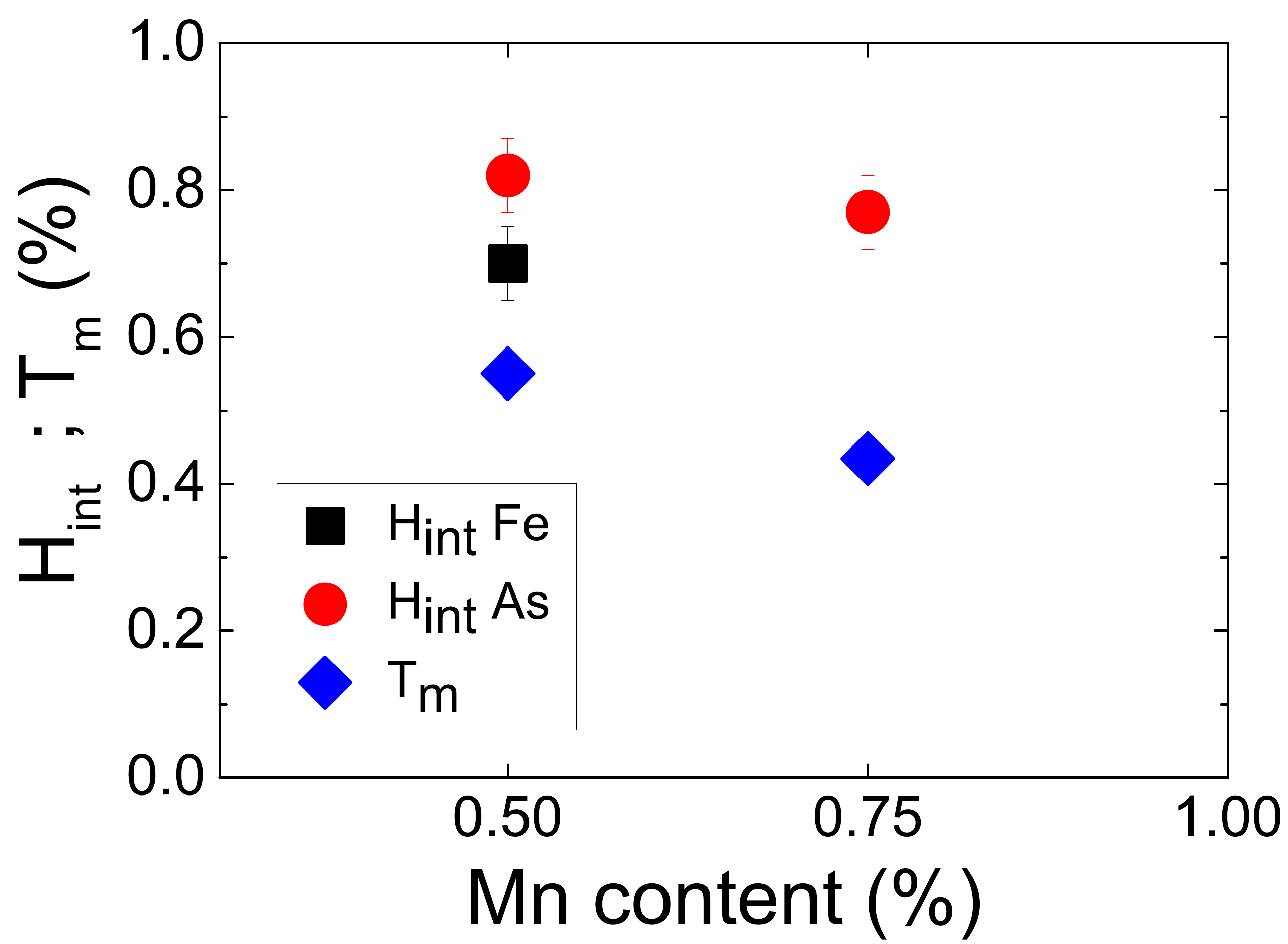} \caption{Intensities of the
			hyperfine magnetic fields at the As (H$_\mathrm{int}$ As) and Fe
			(H$_\mathrm{int}$ Fe) sites at 8 K in
			LaFe$_{1-x}$Mn$_{x}$AsO$_{0.89}$F$_{0.11}$ ($x=$0.5 \% and
			$x=$0.75 \%) normalized to those measured in undoped LaFeAsO. The
			hyperfine fields at the two sites were measured respectively with
			ZF-NMR and \mossbauer spectroscopy. The relative magnetic
			transition temperature, measured with ZF-$\mu$SR, is also
			reported. The reference values measured in LaFeAso are: T$_m=$ 145
			K, H$_\mathrm{int}$ As = 1.6 T and  H$_\mathrm{int}$ Fe = 5 T.}
		\label{fig:fields}
	\end{figure}

	$^{75}$As NQR and ZF-NMR spectra were derived by recording the
	integral of the echo signal after a $\pi/2-\tau_e-\pi$ pulse
	sequence as a function of the irradiation frequency. All the
	ZF-NMR spectra were measured with exactly the same set-up and coil
	filling factor in order to compare the relative intensity of the
	lines. It must also be noted that the length of the pulses was
	optimized for the $m_I=-1/2\rightarrow m_I=-3/2$ line and kept
	constant for the whole spectrum. However the resulting distortion
	in the spectrum amplitude is not sample dependent and does not
	modify the position of the lines.
	
	The resulting spectra both for undoped LaFeAsO and for \lafemnasof
	samples are reported in Fig.~1a of the main part of the paper.
	Since, the stripe magnetic order is also present in \lafemnasof
	(see next section) we can derive the intensity of the hyperfine
	field at the As site from the frequency $\nu_c= (\gamma/2\pi)
	|\mathcal{A}\langle\vec S \rangle|= (\gamma/2\pi)\mathrm{H_{int}}$
	of the low-frequency line corresponding to the $m_I=1/2\rightarrow
	m_I=-1/2$ transition (see main part). The results are reported in
	Fig.~\ref{fig:fields}.

	
	The $^{75}$As NMR experiments were performed by a homemade NMR
	spectrometer and a home-assembled probehead placed in the
	variable-temperature insert of a field-sweeping cold-bore
	cryomagnet. The large capacitance span of the variable capacitor
	in the probehead (approx. 2-100 pF) provides a tuning range of
	more than two octaves. This allowed us to cover the entire
	spectral range of the $^{75}$As and $^{139}$La resonances at 8 T
	with a single coil. The usage of the same coil for all the
	resonances, along with a careful calibration of the frequency
	response of the spectrometer, ensures a reliable quantitative
	comparison of the amplitudes from different spectral features.

	The spectra were recorded by tuning the probehead at discrete
	frequency steps by means of a software-controlled servomechanism
	featured by the spectrometer itself, and exciting a spin echo. The
	spin-echo sequence was a standard $P-\tau-P$ one, with equal rf
	pulses $P$ of duration $\approx 12-16~\mu s$ and intensity
	suitably adjusted to optimize the signal.
	The delays $\tau$ was kept as short as possible with respect to
	the dead time of the resonant probehead ($\approx $20-35$ \mu$s
	depending on the working frequency).

	The fraction of nuclei participating in the majority and minority
	$^{75}$As signals, respectively, was estimated from  the integral
	of the the normalized spectral amplitude (i.e. the amplitude
	divided by the frequency dependent sensitivity $\propto
	\omega^2$). In the case of the minority signal, characterized by a
	hyperfine field estimated in the order of 8-10~T from the NMR
	spectra  in lower external fields (not shown), a further
	correction factor is given by the rf enhancement $\alpha$
	originating from the hyperfine coupling between electronic and
	nuclear spins. Such a coupling, on one hand, amplifies the driving
	rf field at the nucleus, so that the resonance can be excited by a
	rf field reduced by $\alpha$; on the other, it enhances the e.m.f.
	induced in the pick-up by the same factor. In a strong external
	field $B_{ext}$ as in the present case, a factor of the order of
	$\alpha \approx B_{eff}/B_{ext}$ (a formula strictly valid for a
	uniaxial ferromagnet, indeed), where $B_{eff}$ is the effective
	field at the nucleus, is expected, \cite{riedi89} hence $\alpha
	\le 2$. A similar value of $\alpha \approx 2$ is obtained by
	comparing the excitation conditions for the two signals, as the
	minority signal was optimally excited with 6dB extra-attenuation.
	After correcting (i.e. dividing) the minority signal amplitude by
	$\alpha= 2$, we obtained our estimate for the volume fraction of
	the minority estimate $^{75}$As nuclei in the order of 3\% of
	total.
	\begin{figure}[t]
		\includegraphics[width=8.131cm,
		keepaspectratio]{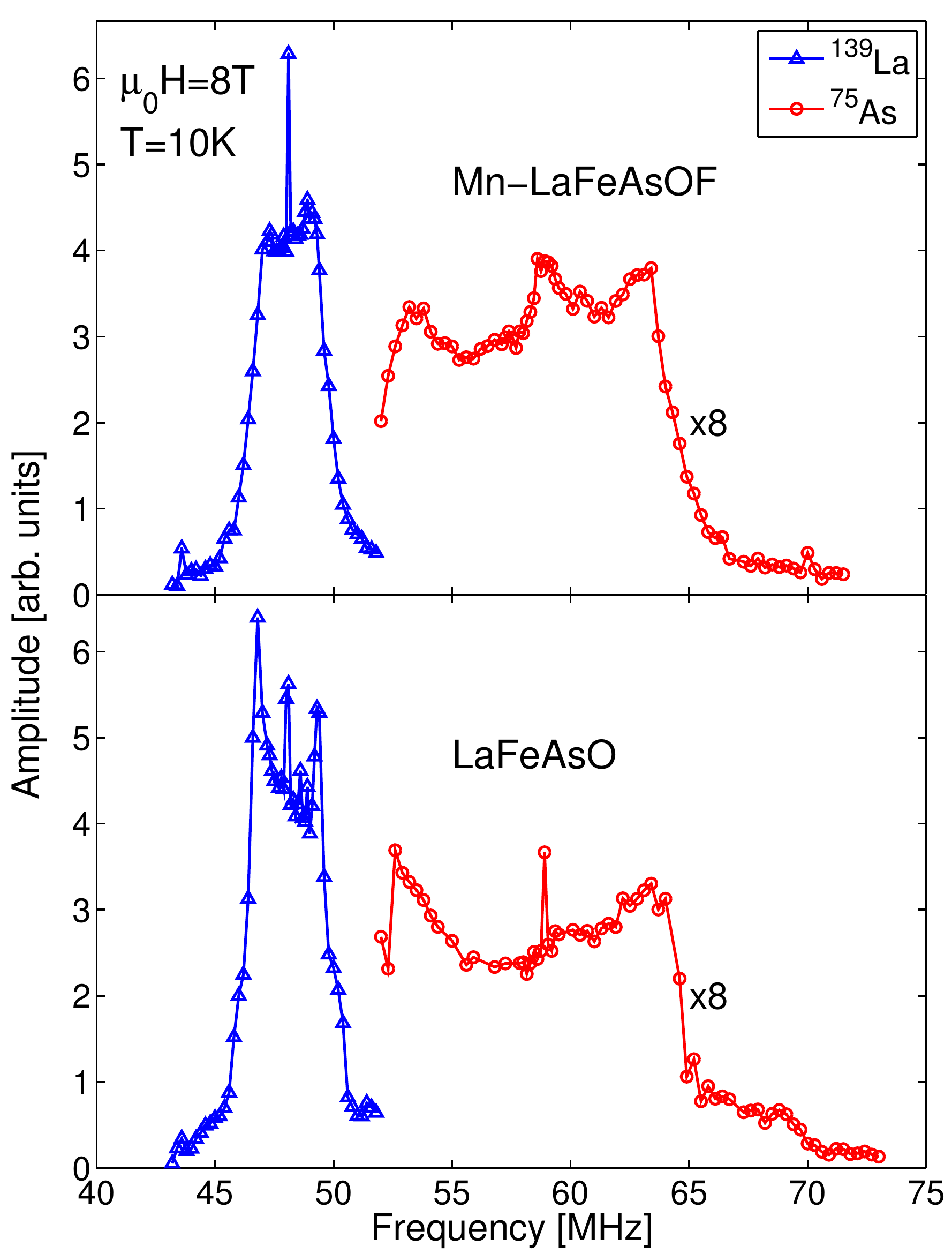} \caption{Top: $^{75}$As (majority signal) and $^{139}$La NMR spectra of LaFe$_{1-x}$Mn$_x$AsO$_{0.89}$F$_{0.11}$ x=0.5\% at $T=10$ K and $\mu_0H_0=8$ T.
			Bottom:  $^{75}$As and $^{139}$La spectra of the parent compound, LaFeAsO, at the same temperature and applied field}
		\label{fig:nmr-spectra-2}
	\end{figure}
	Figure ~\ref{fig:nmr-spectra-2} shows that the As spectrum of the magnetic ordered state in x=0.5\% Mn doped LaFeAsO$_{0.89}$F$_{0.11}$, appearing also in Fig. 2 (main part), is extremely similar to that of the parent compound LaFeAsO, without F doping. This observation provides further evidence that the magnetic orders present in the two compounds is indeed the same.

	\section{IV. Internal field calculations }
	\begin{figure}[t]
		\includegraphics[width=7cm,
		keepaspectratio]{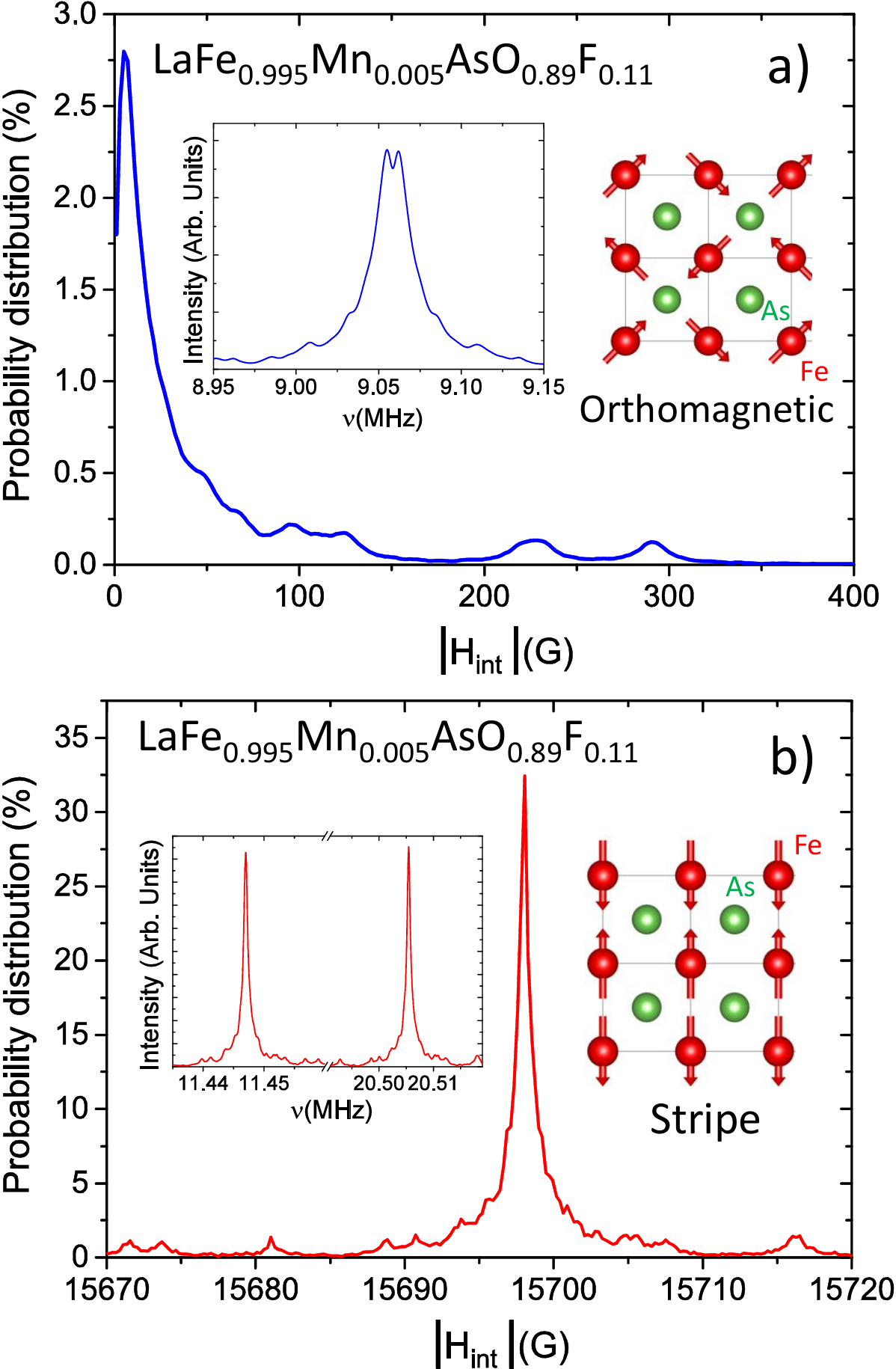} \caption{Simulated (see text) distribution of internal fields on the As site (main panels) and simulated $^{75}$As
			ZF-NMR spectra (left insets) of LaFe$_{0.995}$Mn$_{0.005}$AsO$_{0.89}$F$_{0.11}$
			in case of orthomagnetic order (a) and stripe order (b). In this simulation Mn impurities takes part to the magnetic ordering so their moments have the same orientation of the magnetic moment of the substituted Fe ion, pictorially shown in the right insets.}
		\label{fig:int-field}
	\end{figure}
\begin{figure}[t]
		\includegraphics[width=4cm,
		keepaspectratio]{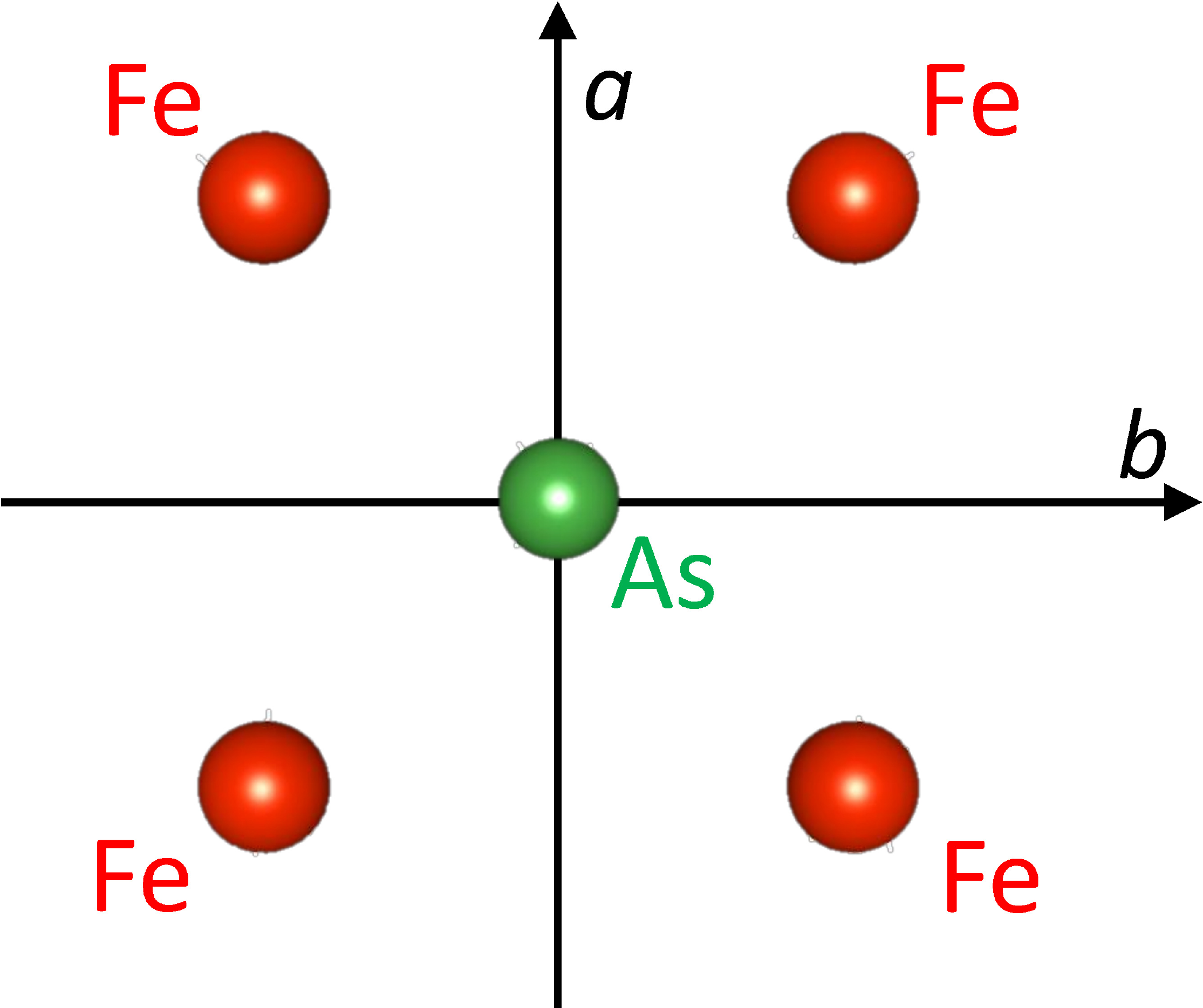} \caption{Sketch of the nearest neighbor Fe sites around an As nucleus in the orthorhombic unit cell. The As does not lie on the same plane of the four Fe sites.}
		\label{fig:nnFe}
	\end{figure}

	In order to interpret the ZF-NMR results we performed simulations
	of the internal field at the As site for different types of long
	range magnetic order and for various Mn concentrations. Both the
	long range dipolar interaction and the short range transferred
	hyperfine interaction between the As nucleus and magnetic moments
	on the four nearest neighbor Fe ions (see Fig.~\ref{hyperfine})
	have been considered in the calculations. The internal field can
	be written as the sum of the contributions from each one of the Fe
	sites:
	\begin{equation} \label{field}
	\textbf{H}_{\mathrm{int}}= \sum_i \textbf{A}_i \cdot \textbf{m}_i
	\end{equation}
	where $\textbf{m}_i$ is the ordered electron moment at the i-th Fe
	site and $\textbf{A}_i$ is the nuclear-electron coupling tensor
	between the As nucleus and i-th Fe site. We considered only the
	contributions due to the Fe sites in the same plane of the As
	nucleus since the contribution to the internal field from the
	other Fe-As layers is vanishingly small due to the $r^{-3}$
	scaling of the dipolar coupling. The diagonal components of the
	symmetric transferred hyperfine interaction tensor for the four
	nearest neighbor Fe sites (Fig.~\ref{fig:nnFe}) was derived from
	Knight shift measurements while two of the three off diagonal
	components can be derived from the strength of the internal field
	in the stripe order configuration, as reported in Ref.
	~\onlinecite{kitagawa2008s}.
	Since we are only interested in understanding which types of order
	give rise to spectra in qualitative agreement with the measured
	ones, we used the values of the transferred hyperfine tensor
	components reported in Ref.~\onlinecite{skitagawa2010s} and chose
	$m_{Fe}= 0.36 ~\mu_B$ for the Fe magnetic moment and $m_{Mn}= 4
	~\mu_B$ for the Mn magnetic moment. The third off-diagonal
	component of the transferred hyperfine coupling is relevant only
	in case of Neel order and was chosen equal to the stripe one. The
	distribution of the internal fields for each type of magnetic
	order was calculated by randomly substituting Mn (for $x=0.5$~\%)
	on the Fe site in a 24~$\times$~24 size mesh and repeating the
	calculation $10^5$ times. The spectra were then obtained by
	diagonalizing the Zeeman-Quadrupole Hamiltonian:
	\begin{align}
		\mathcal{H}&=\mathcal{H}_{\mathrm{Zeeman}}+\mathcal{H}_{Q}=\\\nonumber
		&=-\gamma\hbar\mathrm{\mathbf{H}_{int}}\cdot \mathrm{\mathbf{I}}+ \dfrac{eQV_{ZZ}}{12}[3\mathrm{I}_z^2-\mathrm{I}^2+\eta (\mathrm{I}_x^2-\mathrm{I}_y^2)]
	\end{align}
	for each value of the magnetic field and applying magnetic dipole
	selection rules. The results for orthomagnetic and stripe order
	are reported in Fig.~\ref{fig:int-field}. For the orthomagnetic
	order we found $\mathrm{H_{int}}\sim 0$. This value is
	incompatible with the resonance frequency measured by $^{75}$
	ZF-NMR for x=0.5 \% and 0.75 \% (main part Fig.~1a), which
	displays a sizable internal field with about 80 \% of the value
	found for pure LaFeAsO. In case of N\'eel order the internal field
	is found to be H$_{\mathrm{int}}>1$ T. However in this magnetic
	structure the field is parallel to $a$ and the splitting between
	the NQR lines is expected to be half of the observed one. Another
	possible type of magnetic structure is the spin-charge order
	\cite{giovannetti2011s} which should give rise to two inequivalent
	iron sites. But this is in contrast with the \mossbauer
	measurements which reveal only one iron site. Therefore, the only
	magnetic structure compatible with the observed experimental
	results is the ($\pi$,0) or (0,$\pi$) stripe ordering. In
	addition, one must notice that the linewidth induced by the
	magnetic disorder is three orders of magnitude smaller than the
	one measured. This prediction is in good agreement with the observation that no significant change is induced in the line width by increasing the Mn content from 0.5 \% to 0.75\%. Conversely as can be seen in the NQR spectra (see Ref.~\cite{lang2010s}) fluorine doping has a strong effect on the line width. This implies that the line broadening of \lafemnasof (x=0.5 \% and 0.75 \%) with respect to LaFeAsO is 
	mostly due to the disorder induced by F doping and possibly by the magnetic disorder related to the formation of (0,$\pi$) and ($\pi$,0) magnetic domains that get pinned by the Mn impurities (see Fig.~5 in the main part). Finally it must be noted that we cannot completely rule out the presence of an incommensurate magnetic order with magnetic wave vector very close to the stripe one. In fact the incommensurability leads to a broadening of the ZF-NMR line, which, in this case, would be impossible to observe since it can be much smaller than the line broadening due to fluorine doping.

	\section{V. Model}
	
	A proper modelling of  LaFe$_{1-x}$Mn$_x$AsO$_{0.89}$F$_{0.11}$ includes both a realistic (five-orbital) model of the kinetic energy
	\begin{equation}
	\label{eq:H0}
	\mathcal{H}_{0}=\sum_{\mathbf{ij},\mu\nu,\sigma}t_{\mathbf{ij}}^{\mu\nu}\hat c_{\mathbf{i}\mu\sigma}^{\dagger}\hat c_{\mathbf{j}\nu\sigma}-\mu_0\sum_{\mathbf{i}\mu\sigma}\hat n_{\mathbf{i}\mu\sigma},
	\end{equation}
	with tight-binding parameters determined in
	Ref.~\onlinecite{ikeda10s}, and inclusion of electronic
	interactions given by the multi-orbital Hubbard Hamiltonian
	\begin{align}
		\label{eq:Hint}
		\mathcal{H}_{int}&=U\sum_{\mathbf{i},\mu}\hat n_{\mathbf{i}\mu\uparrow}\hat n_{\mathbf{i}\mu\downarrow}+(U'-\frac{J}{2})\sum_{\mathbf{i},\mu<\nu,\sigma\sigma'}\hat n_{\mathbf{i}\mu\sigma}\hat n_{\mathbf{i}\nu\sigma'}\\\nonumber
		&\quad-2J\sum_{\mathbf{i},\mu<\nu}\vec{S}_{\mathbf{i}\mu}\cdot\vec{S}_{\mathbf{i}\nu}+J'\sum_{\mathbf{i},\mu<\nu,\sigma}\hat c_{\mathbf{i}\mu\sigma}^{\dagger}\hat c_{\mathbf{i}\mu\bar{\sigma}}^{\dagger}\hat c_{\mathbf{i}\nu\bar{\sigma}}\hat c_{\mathbf{i}\nu\sigma},
	\end{align}
	Here $\mu,\nu$ are orbital indexes, ${\mathbf{i}}$ denotes lattice sites, and $\sigma$ is the spin.
	The interaction includes intraorbital (interorbital) repulsion $U$ ($U'$), the Hund's coupling $J$, and the pair hopping energy $J'$. We assume $U'=U-2J$, $J'=J$, and choose $J=U/4$.
	Superconductivity is included by a BCS-like term
	\begin{equation}
	\mathcal{H}_{BCS}=-\sum_{\mathbf{i}\neq \mathbf{j},\mu\nu}[\Delta_{\mathbf{ij}}^{\mu\nu}\hat c_{\mathbf{i}\mu\uparrow}^{\dagger}\hat c_{\mathbf{j}\nu\downarrow}^{\dagger}+\mbox{H.c.}],
	\end{equation}
	with $\Delta_{\mathbf{ij}}^{\mu\nu}=\sum_{\alpha\beta}\Gamma_{\mu\alpha}^{\beta\nu}(\mathbf{r_{ij}})\langle\hat{c}_{\mathbf{j}\beta\downarrow}\hat{c}_{\mathbf{i}\alpha\uparrow}\rangle$ being the superconducting order parameter, and $\Gamma_{\mu\alpha}^{\beta\nu}(\mathbf{r_{ij}})$ denoting the effective pairing strength between sites (orbitals) $\mathbf{i}$ and $\mathbf{j}$ ($\mu$, $\nu$, $\alpha$ and $\beta$). 
	In agreement with a general $s^\pm$ pairing state, we include next-nearest neighbor (NNN) intra-orbital pairing, $\Gamma_{\mu}\equiv\Gamma_{\mu\mu}^{\mu\mu}(\mathbf{r_{nnn}})$. 
	Magnetic disorder modeling the Mn moments is included by $\mathcal{H}_{imp}=I\sum_{\{\mathbf{i^*}\}\mu\sigma}\sigma S_\mu c_{\mathbf{i^*}\mu\sigma}^{\dagger}c_{\mathbf{i^*}\mu\sigma}$, where $S_\mu$ is magnetic moment in orbital $\mu$ at the disorder sites given by the set $\{\mathbf{i^*}\}$ coupled to the spin density of the itinerant electrons. 
	
	After a mean-field decoupling of the interacting Hamiltonian we solve the Bogoliubov-de Gennes equations,
	\begin{eqnarray}
	\begin{pmatrix}
	\hat{\xi}_{\uparrow} & \hat{\Delta}_{\mathbf{ij}}\\
	\hat{\Delta}_{\mathbf{ji}}^{*} & -\hat{\xi}_{\downarrow}^{*} 
	\end{pmatrix}
	\begin{pmatrix}
	u^{n} \\ v^{n} 
	\end{pmatrix}=E_{n}
	\begin{pmatrix}
	u^{n} \\ v^{n} 
	\end{pmatrix},
	\end{eqnarray} 
	where
	\begin{align}
		\hat{\xi}_{\sigma}u_{\mathbf{i}\mu}\!&=\!\sum_{j\nu}t_{\mathbf{ij}}^{\mu\nu}u_{\mathbf{j}\nu}\!+\!\sum_{\mu\neq\nu} [-\mu_0+\sigma I S_{\mu}\delta_{\mathbf{i\{i^*\}}}\delta_{\mu\nu} U n_{\mathbf{i}\mu\overline{\sigma}}\\ \nonumber &\mbox{\hspace{1.9cm}}+U'n_{\mathbf{i}\nu\overline{\sigma}}+(U'-J)n_{\mathbf{i}\nu\sigma} ] u_{\mathbf{i}\mu}\;, \\ \nonumber \mbox{ } \\ \nonumber
		\hat{\Delta}_{\mathbf{ij}}^{\mu\nu}u&_{\mathbf{i}\mu}=-\sum_{\mathbf{j}\nu}\Delta_{\mathbf{ij}}^{\mu\nu}u_{\mathbf{j}\nu}.
	\end{align}
	The five-orbital BdG equations are solved on $30\times30$ lattices with stable solutions found through iterations of the following self-consistency equations 
	\begin{align}
		\label{eq:bdg}
		n_{\mathbf{i}\mu\uparrow} &=\sum_{n}|u_{\mathbf{i}\mu}^{n}|^{2}f(E_{n}),\\\nonumber
		n_{\mathbf{i}\mu\downarrow} &=\sum_{n}|v_{\mathbf{i}\mu}^{n}|^{2}(1\!-\!f(E_{n})),\\\nonumber
		\Delta_{\mathbf{ij}}^{\mu}&=\Gamma_{\mu}\sum_{n}u_{\mathbf{i}\mu}^{n}v_{\mathbf{j}\nu}^{n*}f(E_{n}),
	\end{align} 
	where $\sum_n$ denotes summation over all eigenstates $n$ and $f(E)$ denotes the Fermi function. 
	We stress that the solutions are fully unrestricted and allowed to vary on all lattice sites and orbitals.
	Finally the relative signs of the individual impurity spins $\sigma S_{\mu}$ are obtained by minimizing the total free energy of the system.
	We operate in the regime of interactions where the impurity-free system is paramagnetic ($U<0.9$ eV).
	For the results in Fig.~5, we have used $U=0.87$eV, $kT=0.001$eV, $IS_{\mu}=0.38$eV. 
	A comprehensive description of the band structure and all details of the self-consistent solutions of the mean-field decoupled Hamiltonian in real-space can be found in Ref.~\onlinecite{gastiasoro2016s} and its associated Supp. Material.
	
	Finally we would like to remark that the results clearly show that there is a critical concentration of Mn above which the system leaves the SC phase and is back to the stripe phase which also characterize the undoped LaFeAsO. The critical concentration inferred from the calculation is in good agreement with the experimental one. While the mean distance between Mn impurities is clearly an essential parameter it should be remarked that the recovery of magnetism is a collective and non-local effect so the number of Mn ions is a particular region of the system (e.g. magnetic domains) is not so important (see Fig. 5, main article). For small Mn concentrations no FM/AFM alternation is predicted and anyway such a property is usually not observed in disordered alloys.
	

\end{document}